\documentclass[twocolumn,showpacs,preprintnumbers,amsmath,amssymb,pre,longbibliography]{revtex4-1}
\usepackage{color}    
\usepackage{graphicx}
\usepackage{dcolumn}
\usepackage{bm}
\usepackage{subfigure}
\usepackage{amssymb}
\usepackage{multirow}
\usepackage{amsmath}
\graphicspath{{plots/}}

\renewcommand{\vec}[1]{\mathbf{#1}}

\begin{document}

\title{Energy loss and friction characteristics of electrons at warm dense matter and non-ideal dense plasma conditions}

\author{ Zh.~A.  Moldabekov$^{1}$, T.~Dornheim$^{2}$, M. Bonitz$^{3}$, and T. S.  Ramazanov$^{1}$}

\affiliation{ $^1$Institute for Experimental and Theoretical Physics, Al-Farabi Kazakh National University, 71 Al-Farabi str.,  
  050040 Almaty, Kazakhstan}
  \affiliation{$^2$Center for Advanced Systems Understanding (CASUS), G\"orlitz, Germany}
 \affiliation{$^3$Institut f\"ur Theoretische Physik und Astrophysik, Christian-Albrechts-Universit\"at zu Kiel,
 Leibnizstra{\ss}e 15, 24098 Kiel, Germany}
 % \affiliation{$^{3}$Helmholtz-Zentrum Dresden-Rossendorf, Bautzner Landstra{\ss}e 400, 01328 Dresden, Germany}
\begin{abstract}
We investigate the energy loss characteristics of warm dense matter (WDM) and dense plasmas concentrating on the influence of electronic correlations. The basis for our analysis is a recently developed 
\textit{ab initio} Quantum Monte-Carlo (QMC) based machine-learning representation of the static local field correction (LFC) [Dornheim \textit{et al.}, J. Chem. Phys. \textbf{151}, 194104 (2019)], 
which provides an accurate description of the dynamical density response function of the electron gas at the considered parameters. 
We focus on the polarization-induced stopping power due to free electrons, the friction function, and the straggling rate. In addition, we compute the friction coefficient 
%at WDM and dense plasma conditions, 
which constitutes a key quantity for the adequate Langevin dynamics simulation of ions. Considering typical experimental WDM parameters with partially degenerate electrons, we find that the friction coefficient is of the order of $\gamma/\omega_{pi}=0.01$, where $\omega_{pi}$ is the ionic plasma frequency. 
This analysis is performed by comparing QMC based data to results from the random phase approximation (RPA), the Mermin dielectric function, and the Singwi-Tosi-Land-Sj\"olander (STLS) approximation. 
It is revealed that the widely used relaxation time approximation (Mermin dielectric function)  has severe limitations regarding the description of the energy loss properties of correlated partially degenerate electrons. Moreover, by comparing QMC based data with the results obtained using STLS, we find that energy loss properties are not sensitive to the inaccuracy of the static LFC at large wave numbers $k/k_{F}>2$ (with $k_F$ being the usual Fermi wave number), but  that a correct description of the static LFC at $k/k_{F}\lesssim 1.5$ is important.

% alysed the applicability of the Mermin dielectric function based approximations has been done by comparison of them to the QMC based result.       
% As the result we found the following main results:

\end{abstract}

\pacs{xxx}

\maketitle
\section{Introduction}

The past two decades have witnessed a substantial development in experimental and theoretical research of  high-energy-density plasma physics  \cite{Graziani_2014, Fortov_book, POP2019, bonitz2020,Mabey}. 
Of particular importance is the so-called \emph{warm dense matter} (WDM), which has emerged as a topic of high interest and of active investigation. This is, among other reasons, due to the complicated interplay of thermal excitations, quantum degeneracy effects and Coublomb correlations (see, e.g.,  Ref.~\cite{Dornheim_Phys.Rep.}).   From an astrophysical perspective, WDM exists in the interior of giant planets \cite{Vorberger2007, PhysRevLett.120.115703}, and white as well as brown dwarfs \cite{Becker_2014, Chabrier_2000}. Moreover, WDM as well as partially degenerate dense plasmas are produced in inertial confinement fusion (ICF) experiments \cite{Hu_2010}.

In this context the energy loss of ions in dense plasmas and WDM constitutes one of the key properties for the design of experiments and for the understanding of the time evolution of the plasma \cite{Deutsch, PhysRevLett.122.015002, Sharkov, MRE2018}.  Thus, various methods have been applied to investigate the stopping power, such as molecular dynamics (MD) simulations \cite{Grabowski, Bernstein_2019}, the binary collision approximation \cite{Deutsch, Bernstein_2019, MRE2018}, time dependent density functional theory (TD-DFT) based MD simulations (TD-DFT-MD)~\cite{PhysRevB.98.144302}, nonequilibrium Green functions methods \cite{balzer_prl_18,schluenzen_cpp_18,bonitz_pss_18}, dielectric function approaches \cite{Barriga_POP,Barriga_PRE10, barriga_2008, Zhen-Guo, Bin}, and combinations of various methods, e.g. \cite{Kraeft, Gericke_PRE2009, Zwicknagel}. In contrast to the case of classical plasmas,  an \textit{ab initio} description of the stopping power in WDM  and dense plasmas with partially degenerate non-ideal electrons remains an unsolved problem to this date, as TD-DFT-MD simulations remain computationally expensive and no reliable time dependent exchange-correlation functionals (potentials) exist~\cite{Pribram-Jones, PhysRevB.98.144302, bonitz2020}.

The stopping due to free particles constitutes the dominant mechanism of energy loss in highly ionized media such as dense plasmas and WDM \cite{Deutsch, Zwicknagel}.
In addition, the contribution of ions is important at small velocities (compared to the thermal velocity of the electrons), while at intermediate to large velocities the projectile stopping is mainly due to electrons (see, e.g., Refs~\cite{Deutsch, Zwicknagel}). Because of the latter, the electronic contribution to the stopping power is dominant during the initial heating and compression in indirect ignition scenarios of ICF. 
This particular stage of ICF is characterized by the occurrence of degenerate dense plasmas with correlated electrons, i.e., a path through the WDM regime.  Therefore, for an adequate description of energy deposition in ICF experiments, the computation of the energy loss due to the warm dense electrons is highly needed.  
However, such investigations have hitherto been prevented by the lack of \textit{ab initio} data for the dynamic properties of electrons, in particular the dynamic density response function.

The situation has changed with the recent progress in Quantum Monte Carlo (QMC) simulations of electrons at WDM parameters, e.g., Refs.~\cite{dornheim_jcp_19-nn, groth_prb_19, dornheim_prl_18, bonitz2020,groth_jcp17,dornheim_pre17,dornheim_jcp_19,dornheim_pop17,dornheim_prl16}. 
More specifically, Dornheim, Groth, and co-workers~\cite{dornheim_prl_18,groth_prb_19} have presented the first accurate results for the dynamic density response of the uniform electron gas in the WDM regime based on extensive path integral Monte Carlo data, without any approximation regarding exchange--correlation effects. This was achieved by reconstructing the dynamic local field correction (LFC) $G(k,\omega)$---the key quantity regarding the linear response of correlated electrons. Moreover, the same authors have found that, often (except for low density), (in particular at WDM conditions) the exact static limit $G(k)=G(k,0)$ is sufficient to accurately estimate the dynamic density response~\cite{dornheim_prl_18,groth_prb_19}, and have subsequently presented a highly accurate machine-learning based representation of $G(k)$ covering the entire relevant parameter range~\cite{dornheim_jcp_19-nn}.

In this work, we utilize these new results to carry out a detailed investigation of the energy loss properties of free electrons in the WDM regime. In particular, we focus on the stopping power, straggling rate, and the friction function.

%Therefore analysis of the energy loss properties of free electrons at WDM conditions become possiable. Such investigation is presented in this work. Particularly, we focus on stopping power, straggling rate and friction function. 

We have performed this analysis by comparing QMC based data to widely used models such as the  Singwi-Tosi-Land-Sj\"olander approximation (STLS) \cite{PhysRev.176.589, 1986JPSJ} from dielectric theory and the relaxation time approximation (RTA) \cite{bonitz_qkt}. 
This allows us to gauge the applicability of the RTA (Mermin dielectric function) for the description of the energy loss properties of electrons at WDM conditions. Additionally, comparing QMC results to STLS based data, we investigate the question: what is the requirement for a static LFC model to provide an accurate description of the energy loss properties of an electron gas at the considered WDM and dense plasma parameters.

Another important property is the friction force acting upon an ion due to the surrounding electrons. Although it was recently found that the inclusion of dissipation effects due to friction is critical for an adequate description of the dynamics of non-ideal ions in WDM and in dense  plasmas \cite{PhysRevLett.104.245001, Kang_2018, Mabey,gregori_bohm_18, Hanno_POP}, the friction coefficient at these parameters has not been investigated so far. Instead,  a simplified  expression for the friction coefficient within the Rayleigh model \cite{PhysRevE.77.061136} was used \cite{PhysRevLett.104.245001, Kang_2018, Mabey}. This is potentially worrying, as this approximation  for the friction coefficient does not take into account electronic degeneracy and correlation effects. Therefore, in addition to the stopping power and straggling rate, we present the first results for the friction acting on an ion due partially degenerate and correlated free electrons in this work.

After introducing the dimensionless parameters and briefly covering the underling theory in Secs.~\ref{s:parameters} and ~\ref{s:theory}, we present our new results and the corresponding extended discussion of the stopping power, straggling rate and friction properties in Sec.~\ref{s:results}. In Sec.~\ref{s:friction}, the implications of the obtained results for the Langevin dynamics of ions are discussed. The paper is concluded with a summary of our main finding in Sec.~\ref{s:dis}.

%In Sec.\ref{s:dis} the summary of main findings is given.       

 \section{Dimensionless parameters}\label{s:parameters} 

The electrons in dense plasmas and WDM are characterized by two dimensionless parameters: $r_s=a/a_B$ and $\theta=k_B T_e/E_F$; here $a$ is the mean distance between electrons, $a_B$ is the first Bohr radius, $E_F$ is the Fermi energy, and $n_e$ ($T_e$) is the electronic number density (temperature).
Being classical particles, the ions are characterized by a singe parameter---the coupling parameter $\Gamma=(Ze)^2/(a_i k_B T_i)$, where $Z$ is the ion charge number and $a_i=(4\pi/3~n_i)^{-1/3}$ the mean distance between ions (with the number density $n_i$, and the ion temperature $T_i$). 

For partially and strongly degenerate electrons, the parameter $r_s$ represents a coupling parameter \cite{bonitz2020,ott_epjd18} and defines the density of the electrons via the relation  $n_e\simeq 1.6\times 10^{24}~r_s^{-3}$.  The degeneracy parameter $\theta$ indicates degree of quantum degeneracy~\cite{ott_epjd18} and for partially degenerate electrons it holds $\theta\sim 1$.   
The temperature of the electrons is expressed through the coupling and degeneracy parameters, $r_s$ and $\theta$,  as $T\simeq (\theta/r_s^2)\cdot 0.58\times 10^{6}~{\rm K}$.
% The classical ionic component is described by the coupling parameter $\Gamma=Z_i^2e^2/(ak_BT_i)$, where $a=(3/4 \pi n_i)^{-1/3}$ is the mean inter-ionic distance, $T_i$ is the temperature of ions, and $n_i=n_e/Z_i$.  The temperature of ions and electrons can be different.  A transient non-isothermal state (with $T_i\neq T_e$) appears due to slow electron-ion temperature relaxation rate resulting from the large ion-to-electron mass ratio \cite{Hartley, White2014, MRE2017, Gericke}. 

In this work, we consider correlated electrons at $r_s=1$, $r_s=2$, and $r_s=4$ in the range of degeneracy parameters from $\theta=0.5$ to $\theta=4.0$, i.e., density and temperature in the range $1.6\times 10^{22}~{\rm cm^{-3}}\lesssim  n_e \lesssim  2.5\times 10^{24}~{\rm cm^{-3}}$ and $2\times 10^{4}~{\rm K}\lesssim  T_e\lesssim  2.3\times 10^{6}~{\rm K}$, respectively. 
These are standard experimental parameters for dense plasmas and WDM, see, e.g., Refs.~\cite{Hu_2010, sperling_free-electron_2015, moldabekov_pre_18, Fortov_book}. 

For the energy loss properties, one additional parameter is given by the projectile velocity $v$ divided by a characteristic velocity of the electrons. 
For partially degenerate electrons, it holds $k_B T_e\sim E_F$ and, thus, both the thermal velocity $v_{\rm th}=\sqrt{k_BT_e/m_e}$ and the Fermi velocity $v_F=\hbar k_F/m_e$ are relevant quantities (where $k_F=(3\pi^2 n_e)^{1/3}$).
Therefore, we present the dependence of the energy loss properties on both $v/v_{\rm th}$ and $v/v_F$.

% These values of the parameters are chosen to track the impact of electronic correlations starting from the case of non-ideal electrons at $r_s=1.5$ to the case of weakly coupled electrons at $r_s=0.5$. Additionally, electronic thermal excitation effects can be measured by comparing the strongly degenerate case at $\theta=0.1$ to the semi-classical regime, $\theta=2.0$.    

%  The plasma parameters that are considered in this work have already been realized experimentally
%  during cryogenic DT implosion on OMEGA, direct-drive ignition
%  at the NIF \cite{Hu, Boehly} as well as in other laser-driven shock-compression experiments \cite{Ma, Glenzer, Lee}.  
% % In Ref.\cite{PRE18}, the discussion of the associated plasma parameters are given.  

% Further, for simplicity and without loss of generality, we take $Z_i=1$ and $\Gamma=15$. The corresponding temperature ratio is given by $T_e/T_i\simeq 1.84 \times (\theta/r_s)\Gamma$ and considered to be $T_e/T_i\lesssim 20$ in accordance with experiments (see discussions in Refs.~\cite{PRE18, Clerouin, Harbour, Plagemann}). For example, at $r_s=0.5$ and $\theta=0.1$ we have $T_e/T_i\simeq 5.5$ and at $r_s=1.5$ and $\theta=1$ we find $T_e/T_i\simeq 18$.

\section{Theory} \label{s:theory} 
\subsection{Dynamic density response and dielectric function}

The density response function is an invaluable quantity, which contains the full information on both the static and dynamic properties of electrons, such as the spectrum of excitations  and  the (free) energy~\cite{groth_prl17,PhysRevB.101.045129}.
It is convenient to decompose the full response function $\chi$ into the following form~\cite{Vignale}:
\begin{equation}\label{chi_G}
\chi_e^{-1}(\vec k, \omega)=\chi_0^{-1}(\vec k, \omega)+\frac{4\pi e^2}{k^2}\left[G(\vec k, \omega)-1\right],
\end{equation}
where $\chi_0$ is the ideal density response function of the electrons  and $G$ is the local field correction (LFC), which incorporates all electronic correlation effects. 

We note that the density response function is connected to the dielectric function via the relation:  
\begin{equation}\label{dynamic_epsilon_0}
 \epsilon^{-1}(\mathbf{k},\omega) = 1 +\frac{4\pi e^2}{k^2}\chi_e(\mathbf{k},\omega).
\end{equation}
%
% It  should be noted that, if for $\varphi_{\rm ei}$ instead of the Coulomb potential an electron-ion pseudopotential (such as the so-called empty core potential) is chosen, the effective potential (\ref{eff_pot_1}) must be used.

When electronic correlations are ignored, i.e., $G(\vec k, \omega)=0$, the description reduces to a mean field approximation --
%, which is usually refereed to as 
the random phase approximation (RPA).
Thus, the comparison of the various approaches to RPA results serves as a tool for the evaluation of the importance of electronic correlations (see, e.g., Refs.~\cite{Dornheim_Phys.Rep., moldabekov_pre_18, zhandos_pre_19}), and we do the same in the present work.

Over the years, many approximate models for the LFC have been introduced to go beyond RPA~\cite{PhysRev.176.589,1986JPSJ,tanaka_cpp_2017,Dufty,Scweng,doi:10.1063/1.4969071},
with the approach by Singwi \textit{et al.}~\cite{PhysRev.176.589,1986JPSJ} (STLS) and the RTA~\cite{Mermin1970} (Mermin) being among the most popular ones. Yet, their respective accuracy has remained unclear.
%There are various models of LFC developed to go beyond RPA. 
%STLS and  are standard models for LFC but two.
In the light of the recent \textit{ab inito} data for both the dynamic and the static response function~\cite{dornheim_prl_18,groth_prb_19,dornheim_jcp_19-nn}, it has become possible to check the validity range of these models regarding different physical properties, which we do in this work for the energy loss and related quantities.

%such models used for the investigation of various physical properties of plasmas can be rigorously checked and their applicability range defined.      

\subsection{Density response function and LFC from ab initio simulations}

Until the early 1990s, no accurate data for the LFC had been available, and use was made of partly uncontrolled approximations, that are known to violate different constraints~\cite{Dufty,Dornheim_Phys.Rep.}. This situation has somewhat changed only when Moroni \textit{et al.}~\cite{MCS,MCS2} presented the first reliable data for the static LFC by carrying out ground-state QMC simulations with a harmonic perturbation and subsequently measuring the response. These data (available at $\theta=0$ and $r_s=2, 5$, and $10$) were then used by Corradini \textit{et al.}~\cite{CDOP}, who presented a consistent analytical representation of $G(k;r_s,\theta=0)$ incorporating the correct large- and small-$q$ limits known from theory~\cite{Dufty,Holas}.

Yet, these results were limited to $\theta=0$, and their applicability for WDM conditions remained unclear~\cite{groth_jcp17,dornheim_pre17}. This situation has recently changed, when Dornheim \textit{et al.}~\cite{dornheim_jcp_19-nn} presented a complete, continuous representation of $G(k;r_s,\theta)$ covering the entire relevant parameter range ($0.7\leq r_s \leq 20$, $0\leq\theta\leq4$, and $0\leq k \leq 5k_\textnormal{F}$). This was achieved by combining the Corradini representation in the ground-state limit with extensive new \textit{ab initio} path integral Monte Carlo data for the static LFC at finite temperature to train a fully connected deep neural network (NN). The latter takes as input a tuple of the form $(k/k_\textnormal{F},r_s,\theta)$ and predicts the corresponding LFC with high accuracy within the specified parameter regime. For completeness, we mention that accurate data for the LFC at even lower density, $r_s\geq20$, have recently been presented in Ref.~\cite{PhysRevB.101.045129}.

Although these new data are restricted to the static limit ($\omega=0$), it was shown in Refs.~\cite{dornheim_prl_18,groth_prb_19} that the frequency-dependence of $G(k,\omega)$ has negligible impact on the dynamic density response of the electrons for $r_s\lesssim4$. Thus, the available data for the static LFC are fully sufficient for the present study of the electronic energy loss properties.

%The most reliable data for LFC are from \textit{ab-initio} quantum Monte Carlo simulations.
%Recently, Dornheim et al has development machine learning representation of LFC based on QMC data for entire WDM parameters \cite{dornheim_jcp_19-nn}. 
%It was shown that static LFC provides accurate description of dynamic response function at $r_s=a/a_B\leq 4$. \cite{groth_prb_19, dornheim_prl_18, bonitz2020}. Therefore, in this work we use QMC based machine learning representation of  static LFC \cite{dornheim_jcp_19-nn}. \\
%\textcolor{red}{Tobias, can you please add a brief discussion here to make a good impression :) }.

\subsection{STLS model for the LFC}

The basic assumption of the STLS approach introduced by Singwi \textit{et al.}~\cite{PhysRev.176.589} is that the two-particle distribution function $f(\vec r,\vec p, \vec r^{\prime}, \vec p^{\prime})$ can be approximated by
\begin{eqnarray}
f(\vec r,\vec p, \vec r^{\prime}, \vec p^{\prime})=f(\vec r,\vec p,)f(\vec r^{\prime}, \vec p^{\prime})g(|\vec r-\vec r^{\prime}|) \ ,
\end{eqnarray}
with $g(|\vec r-\vec r^{\prime}|)$ being the radial pair distribution function of the electrons evaluated in thermodynamic equilibrium.
%Based on physically transparent closure for two particle distribution function $f(\vec r,\vec p, \vec r^{\prime}, \vec p^{\prime})=f(\vec r,\vec p,)f(\vec r^{\prime}, \vec p^{\prime})g(|\vec r-\vec r^{\prime}|)$, with $g(|\vec r-\vec r^{\prime}|)$ being radial pair distribution function of electrons, STLS scheme was widely used to approximate the static LFC for computation of various properties of dense plasmas.
In particular, STLS-based methods were extensively applied to study the energy loss in dense plasmas and WDM~\cite{Zwicknagel, Montanari,Gauthier, Wang, Barriga, Barriga_PRE10}.

% as well as other dynamic properties like transport and relaxation   charactersitics \cite{Bennadji, Reinholz95, Benedict} of .  

The STLS approximation is defined by the following equation for the static LFC:
\begin{equation}
G^\textnormal{STLS}(\mathbf{k},0) = -\frac{1}{n} \int\frac{\textnormal{d}\mathbf{k}^\prime}{(2\pi)^3}
\frac{\mathbf{k}\cdot\mathbf{k}^\prime}{k^{\prime 2}} [S^{\text{STLS}}(\mathbf{k}-\mathbf{k}^\prime)-1]\;,
\label{G_stls}
\end{equation}
where the fluctuation-dissipation theorem is used as the closure relation for the static structure factor $S^{\textnormal{STLS}}$, 
\begin{equation}\label{flucDis}
 S^\text{STLS}(\mathbf{k}) = -\frac{1}{\beta n}\sum_{l=-\infty}^{\infty} \frac{k^2}{4\pi e^2}\left(\frac{1}{\epsilon(\mathbf{k},z_l)}-1\right)\ .
\end{equation}
The inverse dielectric function is then computed via Eq.~(\ref{dynamic_epsilon_0}) using 
%the finite temperature ideal response function 
$\chi_0$ and $G^\text{STLS}$.  Further, we mention that the summation in Eq.~(\ref{flucDis}) is typically carried out in imaginary-time, leading to the Matsubara frequencies, $z_l=2\pi il/\beta\hbar$.

\subsection{Relaxation time approximation (RTA)}
%
%Probably the first and 
The RTA is among the most widely used approximations to account for inter-particle correlations. 
From a quantum kinetic theory perspective, it is based on the introduction of a constant relaxation time $\tau = 1/\nu$ for the transition of the quantum distribution function to an equilibrium state \cite{bonitz_qkt, 1742-6596-220-1-012003}.  

The  RTA (Mermin) quantum dielectric function~\cite{Mermin1970} reads: 
 \begin{eqnarray}\label{Mermin}
 &\epsilon&_{M}(\vec{k},\omega)= \nonumber \\ & & 1+\frac{(\omega+i\nu)[\epsilon_{\text{RPA}}(\vec{k},\omega+i\nu)-1]}{\omega+i\nu[\epsilon_{\text{RPA}}(\vec{k},\omega+i\nu)-1]/[\epsilon_{\text{RPA}}(\vec{k},0)-1]},
 \end{eqnarray}
where $\nu$ denotes the apriori unknown electron collision frequency. In general, both electron-electron  and electron-ion collisions can be included into $\nu$.
In this work, we study the energy loss properties of an electron gas and, therefore, consider only the contribution due to electron-electron collisions,  $\nu=\nu_{\rm ee}$. 
Among other applications, the Mermin dielectric function was often used to compute the stopping power of dense plasmas and to analyze the impact of correlation effects \cite{Barriga_POP, barriga_2008, Barriga, Barriga_PRE10}.  

% In general, both electron-electron  and electron-ion collisions can be included into $\nu$. In this work, to keep analysis and discussion clear, we assume that  $\nu$ is only due to electron-electron collisions. 

% can with $\nu_{\rm ei}$ and $\nu_{\rm ee}$ being the contributions due to , respectively. 
% In Eq.~(\ref{Mermin}), $\epsilon_{\text{RPA}}(\vec{k},\omega)$ is the Lindhard dielectric function \cite{Lindhard}, i.e. the dielectric function in the random phase approximation (RPA).
% Going beyond the static relaxation time approximation, dynamical electronic collision effects are included in the Mermin dielectric function~\cite{Reinholz2000} 

\subsection{Energy loss properties}
The stopping power represents the mean energy loss of a projectile, $\delta E$, per unit path length $\delta l$, i.e. $S=\delta E/\delta l$. 
The standard expression for the stopping power acting upon an ion due to the polarization of the surrounding electronic medium in plasmas reads \cite{PhysRevA.23.1898}: 
\begin{equation}\label{eq:sp}
    S(v)=\frac{2Z^2e^2}{\pi v^2}\int_0^{\infty} \, \frac{{\rm d}k}{k} \, \int_0^{kv} {\rm d} \omega ~\omega ~{\rm Im} \left[\frac{-1}{\epsilon(k,\omega)}\right],
\end{equation}
where $Ze$ is the ion charge and $v$ is the ion velocity.  Eq.~(\ref{eq:sp}) constitutes the linear response result, which is applicable for low-Z projectiles, i.e., when the ion-electron coupling is weak \cite{Zwicknagel}. 
% Additionally, Eq.~(\cite{eq:sp}) 

At small projectile velocities $v/v_{\rm th}\ll 1$, the stopping power due to free electrons~[cf.~Eq.~(\ref{eq:sp})] is linearly proportional to the projectile velocity. 
This is the motivation to introduce the so-called friction function $Q(v)$, which is defined by the relation \cite{Montanari}:
\begin{equation}\label{eq:fric}
    S(v)= Z^2e^2 Q(v)\cdot v\,,
    \end{equation}
which is a constant at low $v$.
As mentioned in the Introduction, an important  quantity for the simulation of non-ideal ion dynamics in WDM and in dense  plasmas is the friction coefficient, which can be computed if the friction function is known. We discuss the friction force acting on ions in Sec.~\ref{s:friction} after presenting results for the friction function.  

Another energy loss property of interest is the energy-loss straggling rate, $\Omega$. The straggling rate describes the statistical fluctuations of the stopping power~\cite{PhysRevA.23.1898} and can also be understood as the variance of the energy loss per unit path length \cite{Barriga_POP},  
\begin{align}
\Omega^2=\frac{\langle (\delta E - \langle\delta E\rangle)^2\rangle}{\delta l}.    
\label{eq:straggling}
\end{align}
Therefore, the straggling rate characterizes the variance of the energy loss of the projectile in a target and is, thus, related to the statistical distribution of the penetration depth of the projectiles as well as of the stopping range (see, e.g., Refs.~\cite{Barriga_POP, PhysRevA.23.1898}).   
% Therefore, straggling rate is important quantity for computation of  
%
Within linear response theory, 
%Linear response 
the result for the struggling rate on the same level of approximation as Eq.~(\ref{eq:sp}) reads \cite{PhysRevA.23.1898, Barriga_POP}:
\begin{equation}\label{eq:strag}
    \Omega^2=\frac{2Z^2e^2\hbar}{\pi v^2}\int\limits_0^{\infty} \frac{{\rm d}k}{k} \int\limits_0^{kv} {\rm d} \omega ~\omega^2\left[ 2N(\omega)+1\right]{\rm Im}\left[\frac{-1}{\epsilon(k,\omega)}\right],
\end{equation}
where $N(\omega)=\left(\exp(\hbar \omega/k_BT_e)-1\right)^{-1}$ is the Planck (Bose) function.

In the following, the results computed using the static LFC from the QMC based machine-learning representation~\cite{dornheim_jcp_19-nn} will be referred to as QMC results. 
Correspondingly, similar short references will be used for the RPA, STLS and Mermin dielectric function-based results.

\section{Numerical results and discussion}\label{s:results} 

\begin{figure}
% [h!]
  \vspace{0.45cm}
\includegraphics[width=0.45\textwidth]{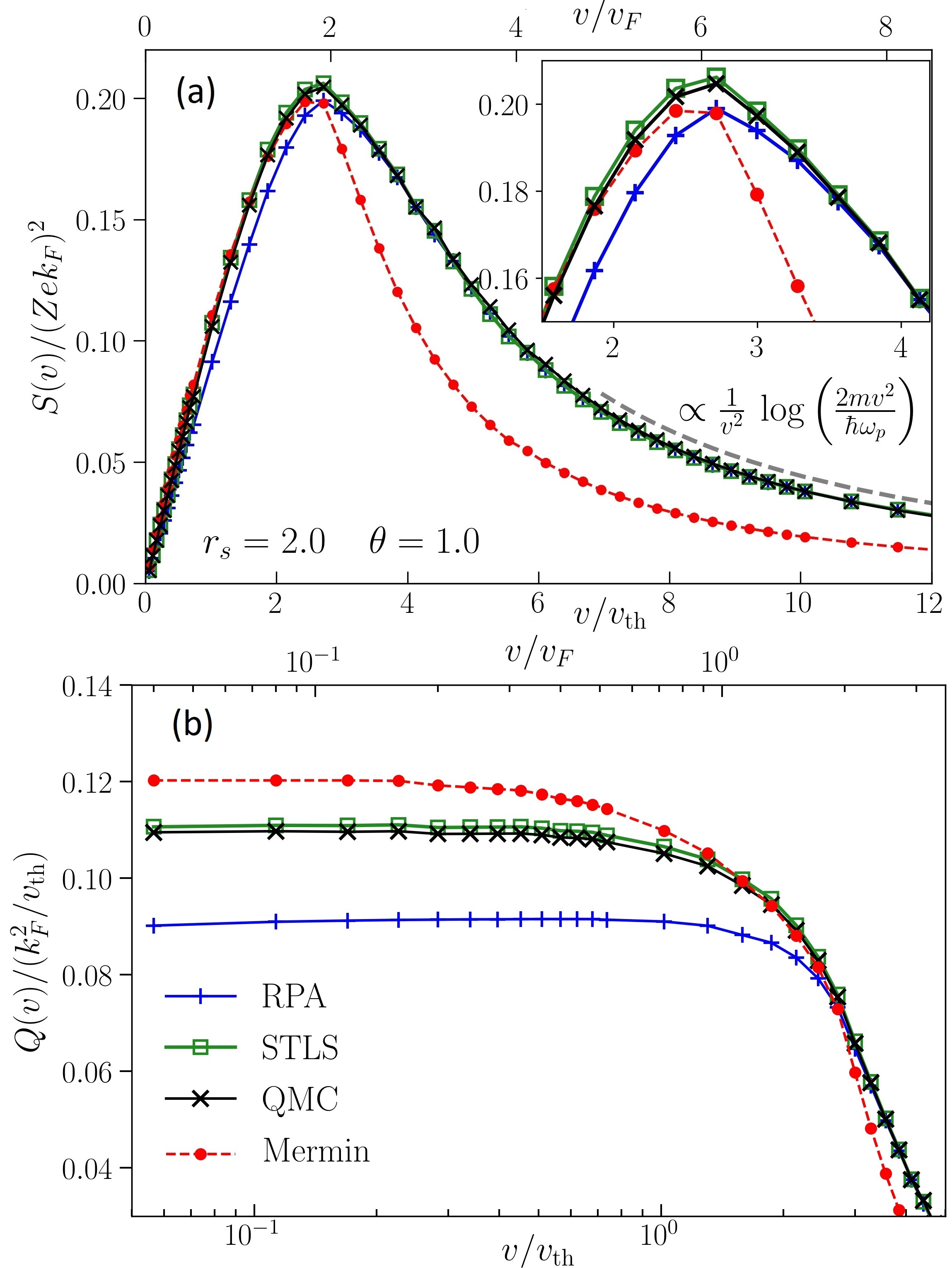}
\caption{(a) Stopping power and (b) friction function at $r_s=2$ and $\theta=1$. The data were obtained using QMC and STLS static local field corrections, the Mermin dielectric function (with $\nu=0.22\omega_p$), and RPA.   
}
\label{fig:1}
\end{figure}
\subsection{Stopping power and friction function}
Let us start our investigation at $r_s=2$ and $\theta=1$ which are typical WDM conditions, where both thermal effects and electronic correlations are important. In Fig.~\ref{fig:1},  we show the stopping power and straggling rate, where the lower x-axis corresponds to $v/v_{\rm th}$ and the upper x-axis to $v/v_{F}$.
It is customary and natural to express the electronic stopping power $S$ in units of $(Zek_F)^2$, so that $S/(Zek_F)^2$ does not depend on the ion charge. Using $S/(Zek_F)^2$ and the thermal velocity, we express the friction function in dimensionless form as $Q/(k_F^2/v_{\rm th})$.

At $r_s=2$ and $\theta=1$, we set for the electron-electron collision frequency $\nu_{\rm ee}/\omega_p=0.22$ in the Mermin dielectric function, with $\omega_p$ being the usual plasma frequency of the electrons. In particular, this value of $\nu_{\rm ee}$ reproduces the QMC static local field correction based result at intermediate velocities $v\approx v_{\rm th}$ [cf.~Fig.~\ref{fig:1}(a)] and is in accordance with a model for the electron-electron collision frequency developed for the computation of the transport properties of quantum plasmas in neutron stars and white dwarfs \cite{PhysRevD.74.043004, Potekhin1999}, which also was used to study the stopping power and straggling rate by Barriga-Carasco \textit{et al.}~\cite{Barriga_POP, barriga-carrasco_potekhin_2006}.
We note that there exists an abundant literature on the models for the collision frequency (e.g., see Refs~\cite{Daligault_prl, Daligault_pop2016, Lee-More, PhysRevD.74.043004, Potekhin1999}). While a detailed analysis of $\nu$ is beyond the scope of the present work, the comparison of the Mermin results to the QMC data that can be considered benchmarks, allows us to draw important general conclusions about the RTA below.

First of all,  by comparing the RPA and QMC results in Fig.~\ref{fig:1}(a), we see that at $r_s=2$ and $\theta=1$ the effects of electronic correlations are important at $v/v_{\rm th}<3$ ($v/v_{F}<2$) and lead to larger values of the stopping power, but the velocity corresponding to the maximum of the stopping power remains almost unchanged. 
At higher velocities, $v/v_{\rm th}>3$ ($v/v_{F}>2$), the difference between QMC and RPA results vanishes.
Secondly, from Fig.~\ref{fig:1}(a) we observe that the STLS result is in remarkably good agreement with the QMC data. This point is discussed below.
%We return to the discussion of this point later. 
The third  observation is that the Mermin dielectric function-based result (with $\nu_{\rm ee}/\omega_p=0.22$) is in good agreement with the QMC result, up to $v=2v_{\rm th}$. At larger velocities, the Mermin result significantly underestimates the stopping power compared to both QMC and RPA. At $\theta\sim1$, such a behavior of the RTA result, as compared to the RPA, is typical and was also reported in previous studies, see Refs.~\cite{PhysRevE.76.016405, barriga-carrasco_potekhin_2006}. 
Moreover, as shown in Fig.~\ref{fig:1}(a), we note that the computed QMC, STLS, and RPA results obey the correct Bethe limit at large velocities, i.e., $S\to (Ze\omega_p/v)^2 \log[2m_ev^2/\hbar \omega_p]$, at $v\to\infty$ \cite{Zwicknagel}. 

At small values of the projectile velocity, the stopping power decreases, and so does the difference between different models. A better picture about the impact of electronic correlations at small velocities 
is revealed by considering the friction function. Just as for the stopping power, the comparison of the STLS based result to QMC data in Fig.~\ref{fig:1}(b) shows very good agreement between them.
From Fig.~\ref{fig:1}(b) we see that, at $v\ll v_{F}$, the friction function is constant, which is similar to the ground-state limit~\cite{PhysRevB.16.115}.
At $r_s=2$ and $\theta=1$, the QMC based result shows that the friction function is constant at $v/v_F<0.5$ ($v/v_{\rm th}<0.8$). Note that the RPA result goes to a constant value with decreasing $v$, starting from larger velocity, $v/v_F<0.9$ ($v/v_{\rm th}<1.5$). At $r_s=2$ and $\theta=1$, electronic correlations taken into account using the QMC LFC lead to a $\sim 22\%$ increase in the friction function, as compared to the RPA.  Note that for the maximum of the stopping power, at $v/v_{\rm th}\simeq 2.7$ ($v/v_{F}\simeq 1.9$), the increase due to electronic correlations is only $\sim 5\%$, and at $1\leq v/v_{\rm th}\leq 2$ it is about $10\%$. 
Therefore, the friction function constitutes a highly sensitive measure to electronic non-ideality effects.
%is much more sensitive quantity to an electronic non-idelaity.

  \begin{figure}[h!]
  \vspace{0.6cm}
\includegraphics[width=0.45\textwidth]{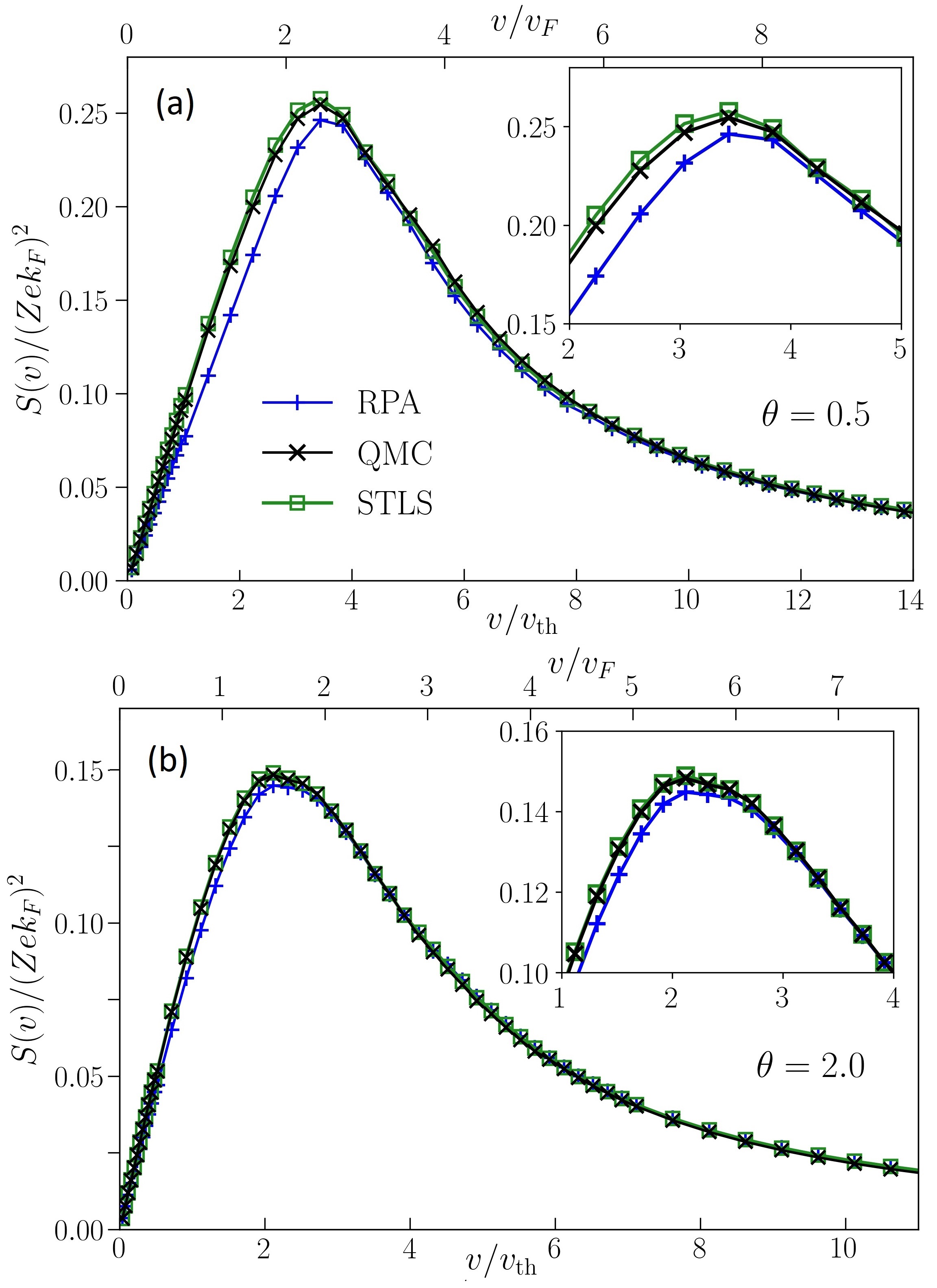}
\caption{Stopping power for $r_s=2$ at (a) $\theta=0.5$ and (b) $\theta=2.0$. The data were obtained using QMC, STLS, and RPA.}
\label{fig:2}
\end{figure}
%----------------------

  \begin{figure}[h!]
 % \vspace{0.6cm}
\includegraphics[width=0.45\textwidth]{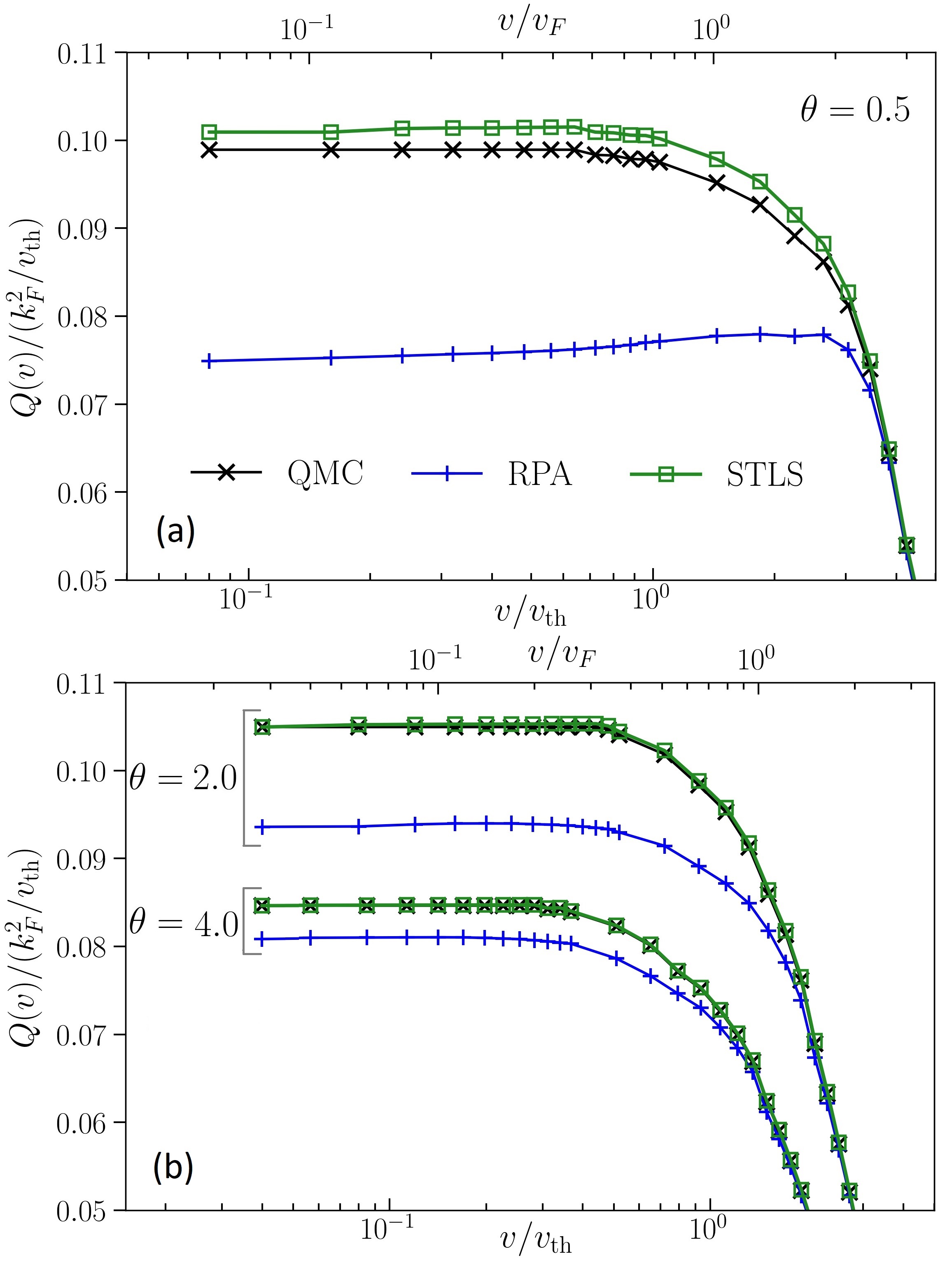}
\caption{Friction function at $r_s=2$ is presented for (a) $\theta=0.5$, and (b) $\theta=2.0$ and $\theta=4.0$.}
\label{fig:3}
\end{figure}

  \begin{figure}[h!]
 % \vspace{0.6cm}
\includegraphics[width=0.45\textwidth]{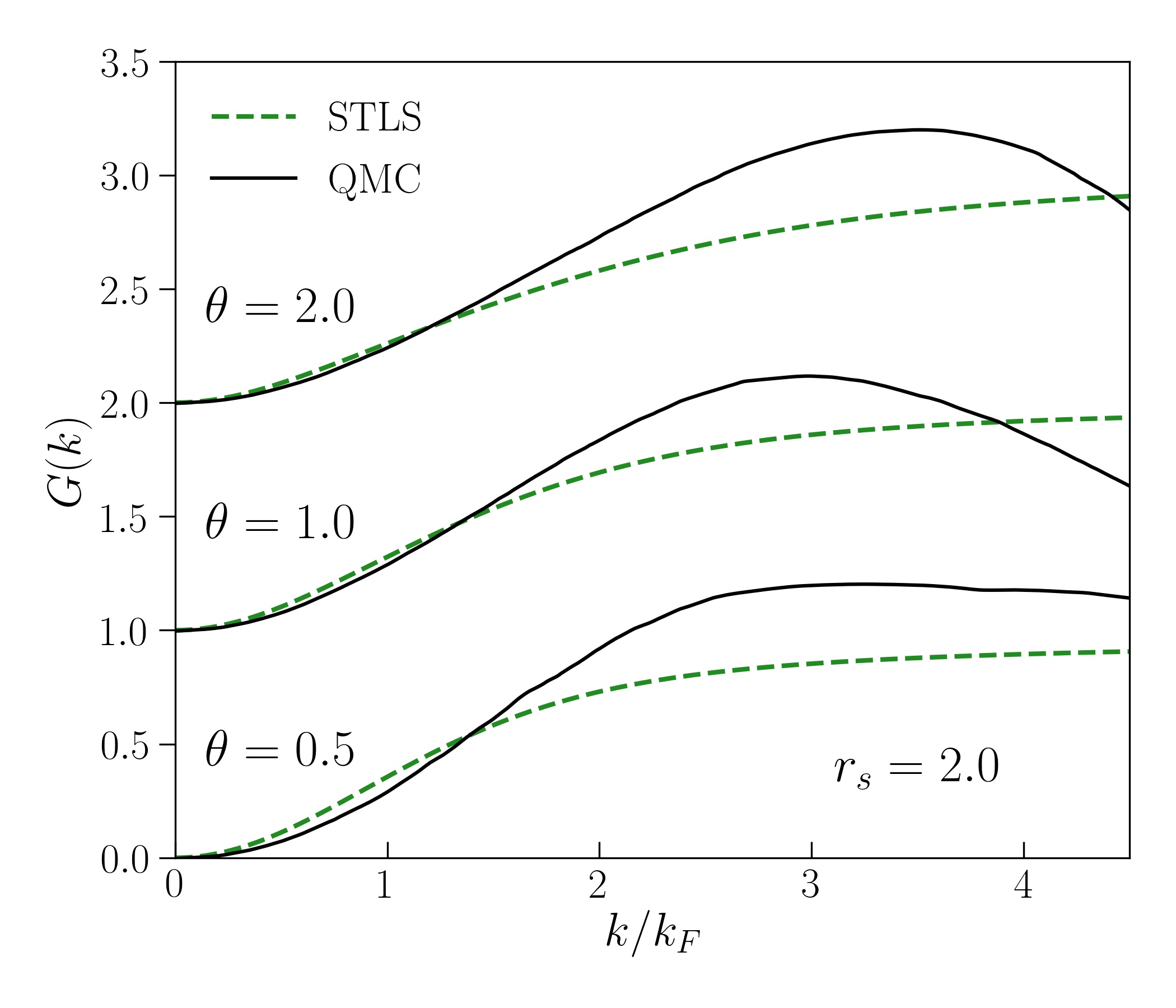}
\caption{Static local field correction at $r_s=2$ and different $\theta$. The STLS result and QMC based machine learning representation \cite{dornheim_jcp_19-nn} are shown.}
\label{fig:4}
\end{figure}

From Fig.~\ref{fig:1}(b), one can see that the relaxation time approximation (with $\nu_{ee}/\omega_p=0.22$) becomes constant at smaller velocities $v/v_{\rm th}<0.4$ ($v/v_F<0.5$) compared to both QMC and RPA results. Moreover, despite a good agreement with QMC result for the stopping power at $v\approx v_{\rm th}$, the Mermin dielectric function based result overestimates the value of the friction function by $9\%$ compared to QMC based result at $v/v_{\rm th}<0.4$. 
Essentially, a straightforward check by the variation of the collision frequency shows that the collision frequency cannot be adjusted in a way that the Mermin result comes to agreement with the QMC based data at all considered velocities for both stopping power and friction function. We thus conclude that the relaxation time approximation has severe limitations regarding the description of the stopping power and friction function at WDM conditions.

Now we extend our analysis of the QMC based results at $r_s=2$ by comparing them to STLS and RPA results at different values of the degeneracy parameter $\theta$. 
To this end, we show the stopping power at $\theta=0.5$ and $\theta=2$ in Fig.~\ref{fig:2}. It can be clearly seen that at both temperatures the STLS results for the stopping power and friction function remain in good agreement  with the QMC curve. Similar to $\theta=1.0$ case, the electronic correlations do not significantly affect the position of the stopping power maximum at $\theta=0.5$ and $\theta=2$ .
Going to higher temperatures at a constant density decreases correlation effects, and the difference between RPA and  QMC (STLS) results at small to intermediate values of the velocity decreases (compare Fig.~\ref{fig:2}(b) to Fig.~\ref{fig:2}(a) and Fig.~\ref{fig:1}(a)). 
Consequently, we find deviations in the maximum stopping power in RPA of $\sim2\%$ at $\theta=2$, and it even vanishes at $\theta=4$.

%At $\theta=2$, the difference between the QMC and RPA results in the value of the stopping power at its maximum is $\sim 2\%$ and vanishes at $\theta=4$.

%In Fig.~\ref{fig:3}(a), the friction function at $\theta=0.5$ and in Fig.~\ref{fig:3}(b) the friction function at $\theta=2$ and $\theta=4$ is presented.
In Fig.~\ref{fig:3}, we show the friction function at $\theta=0.5$ (a) and $\theta=2,4$ (b).
At the lowest depicted temperature, the STLS static LFC slightly overestimates the small velocity limit of the friction function by $3\%$ compared to QMC based data. At $\theta=2$ and $\theta=4$, the difference between STLS results and QMC results is negligible.
With decreasing temperature, the difference between the QMC and RPA results for the friction function in the small velocity limit $v\ll v_{\rm th}$ increases from $22\%$ for $\theta =1$ to $33\%$ for $\theta=0.5$. Moreover, the difference between QMC and RPA results at $v\ll v_{\rm th}$ reduces to $10\%$ at $\theta=2$ and to $6\%$ at $\theta=4$. 
In addition, taking into account correlations, the maximal value of the velocity up to which the friction function is constant reduces from $v/v_{\rm th}=1$ at $\theta=0.5$ to $v/v_{\rm th}=0.4$ at $\theta=4.0$. The same behavior also holds for the RPA result.  

 \begin{figure}[h!]
 % \vspace{0.6cm}
\includegraphics[width=0.45\textwidth]{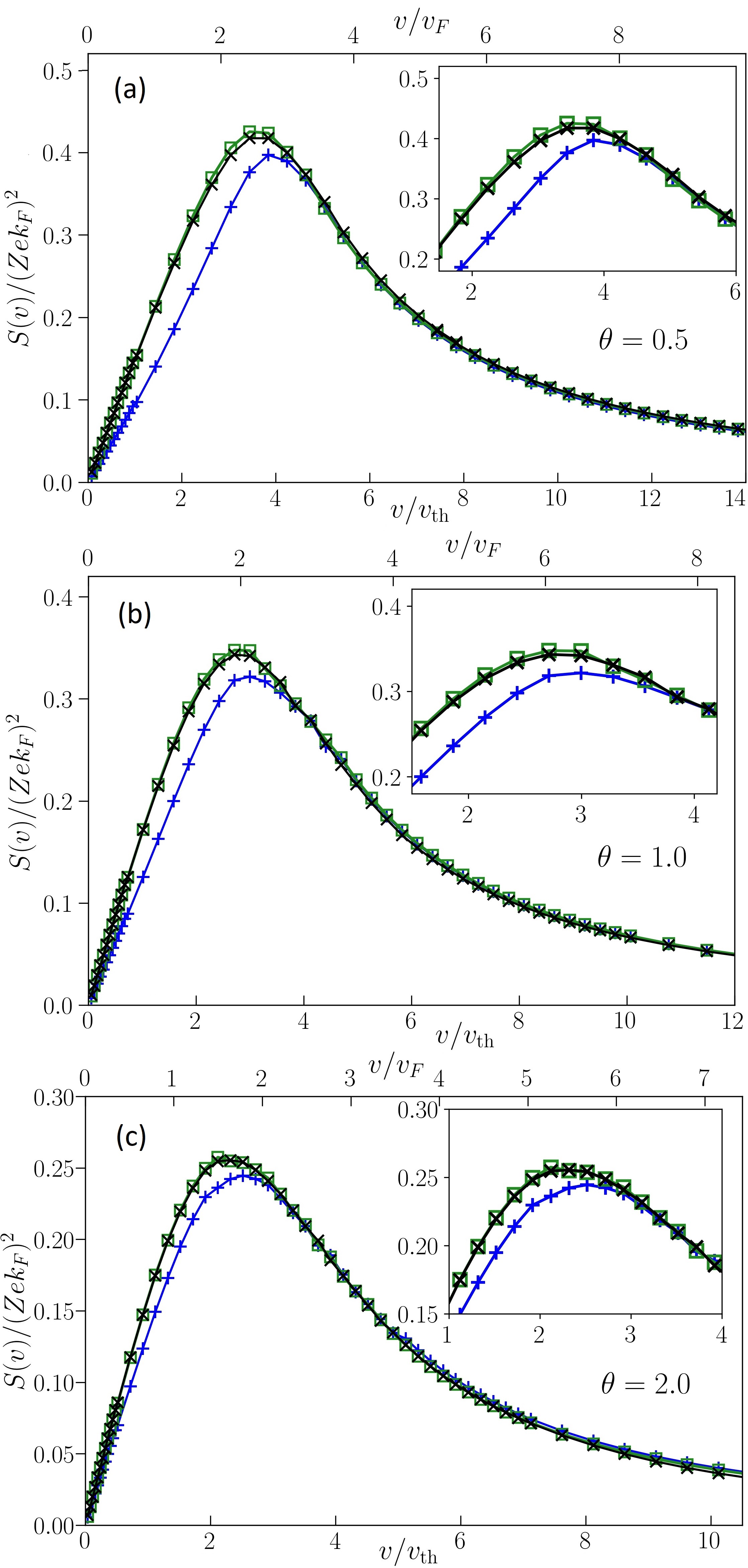}
\caption{Stopping power at $r_s=4$. Results are presented for (a) $\theta=0.5$, (b) $\theta=1.0$, and (c) $\theta=2.0$.}
\label{fig:5}
\end{figure}

To understand the reason for the remarkable agreement between the STLS and QMC results at $r_s=2$, we compare the static LFC from the QMC based machine learning representation \cite{dornheim_prl_18} to the STLS static LFC in Fig.~\ref{fig:4} for different values of the degeneracy temperature $\theta$. We find that the STLS static LFC attains values that are close to the QMC data for wave numbers $k\leq 1.5~k_F$, but exhibits significant deviations for $k > 1.5~k_F$. 
Therefore, taking into account the previously observed agreement between STLS and QMC for the stopping power and friction function we conclude that \textit{the stopping power and related friction function are not sensitive to the LFC values at $k>1.5~k_F$}. This is due to the pre-factor $4\pi/k^2$ in Eq.~(\ref{chi_G}), which suppresses the effect of LFC at large $k$. 
To further support this finding, we next consider $r_s=4$, where electronic correlation effects are even more important.
%To support that this is indeed the case, next we consider $r_s=4$. 

  \begin{figure}[h!]
 % \vspace{0.6cm}
\includegraphics[width=0.45\textwidth]{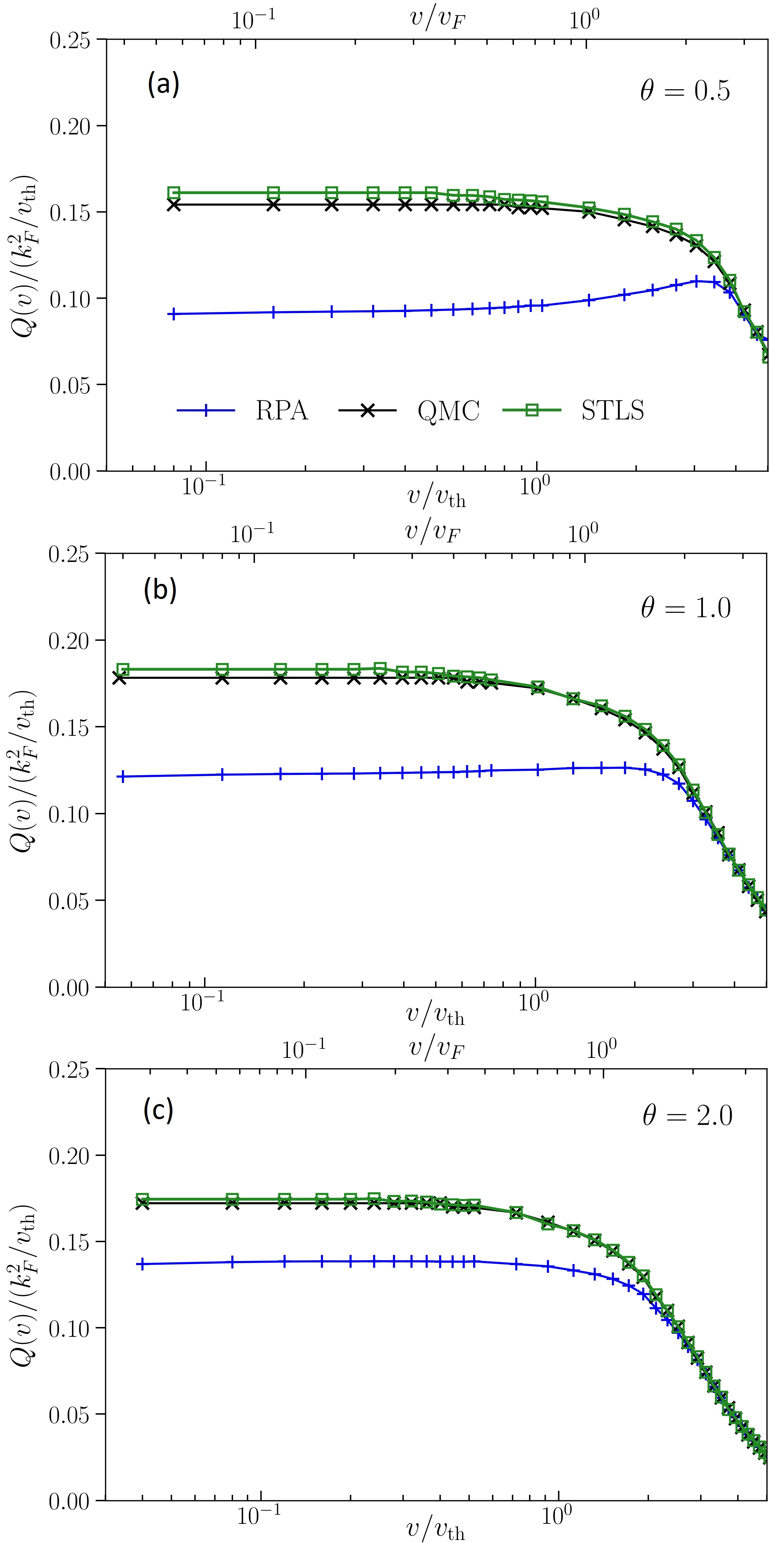}
\caption{Friction function at $r_s=4$. Results are presented for (a) $\theta=0.5$, (b) $\theta=1.0$, and (c) $\theta=2.0$.}
\label{fig:6}
\end{figure}

\begin{figure}[h!]
 % \vspace{0.6cm}
\includegraphics[width=0.45\textwidth]{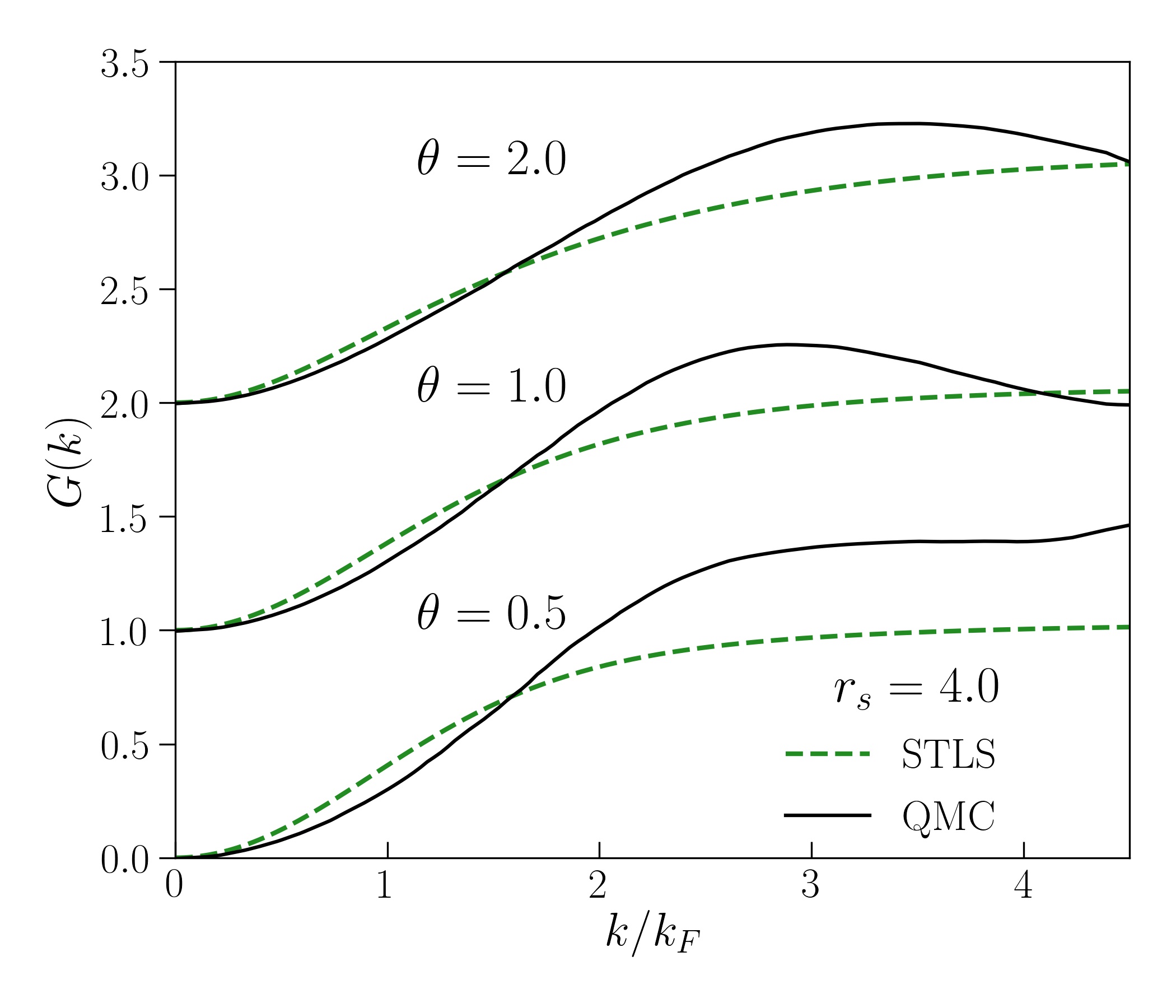}
\caption{Static local field correction at $r_s=4$ and different $\theta$. The STLS result and QMC based machine learning representation \cite{dornheim_jcp_19-nn} are shown. }
\label{fig:7}
\end{figure}

In Figs.~\ref{fig:5} and \ref{fig:6}, we show the stopping power and friction function at $r_s=4$ for three values of $\theta$. 
%At different values of the degeneracy parameter and $r_s=4$, the stopping power and friction function are presented in Figs.~\ref{fig:5} and ~\ref{fig:6}, respectively.
Remarkably, here, too, we observe that the STLS result is in very good agreement with QMC based data.
The corresponding data for the static local field correction at the same conditions are shown in Fig.~\ref{fig:7}. Again, we find that STLS constitutes a decent approximation to the exact LFC for intermediate $k$, where the impact of the LFC on observable properties is most pronounced, whereas the striking deviations at $k\gtrsim 2k_\textnormal{F}$ are of almost no consequence. 

% \textcolor{red}{TD: Comment: at $k\to 0$, the STLS LFC is not exact, but the LFC has no impact, as even RPA is exact (perfect screening, ...). at large $k$, the LFC, too is of no consequence due to the $4\pi/k^2$ pre-factor in Eq.~(\ref{chi_G}). Thus, STLS gives, overall, a rather shitty description of $G(k)$ itself, but this still results in relatively good observables like $\chi(k)$, $S(k)$. This is not a coincidence, as the closure relation for STLS involves $S(k)$ and is, in some sense, "designed" to give good observables, whereas "unimportant" regions of $G(k)$ are not optimized. I don't know, if we want to include this discussion somewhere. }

%From comparison of the corresponding static local field corrections (see Fig~\ref{fig:7}), we see that the reason for that is the close values of the STLS LFC to QMC LFC at small wavenumbers $k\lesssim 1.5~k_F$. 

Comparing the QMC results to the RPA results, we see from Figs.~\ref{fig:5} and ~\ref{fig:6} that the increase of the electronic coupling parameter from $r_s=2$ to $r_s=4$ leads to a further increase of the effect of electronic correlations, and the quality of the mean field approach deteriorates. At $1<v/v_{\rm th}<3$ and $r_s=4$, 
the electronic correlations cause an increase of the stopping power by about $25\%$ at $\theta=0.5$ and $\theta=1.0$. This number decreases to about $10\%$ for $\theta=2.0$. 

The friction function is affected by the inclusion of the electronic non-ideality more strongly than the stopping power, similar to the $r_s=2$ case. 
At $\theta=0.5$, electronic correlations result in a significant ($78\%$) increase of the friction function in the small velocity limit. 
Upon increasing the degeneracy parameter to $\theta=1.0$ ($\theta=2.0$) this number drops to $46\%$ ($20\%$), but clearly remains significant. 
From the QMC results for $r_s=4$ shown in Fig.~\ref{fig:6}, we see that, at $0.5\leq \theta\leq 4$, the friction function can safely be considered as constant at velocities $v/v_{\rm th}\leq 1$ .

Let us now turn to the weaker coupling  with $r_s=1$ to further analyse the impact of the static LFC on the energy loss characteristics and, in this way, to gauge the importance of electronic non-ideality effects. We note that this case is particularly interesting as $r_s=1$ approximately constitutes the boundary between the ideal and non-ideal regimes for partially (or strongly) degenerate electrons.

% \subsection{Yukawa model with a fitting parameter}\label{s:4B}

  \begin{figure}[h!]
 % \vspace{0.6cm}
\includegraphics[width=0.45\textwidth]{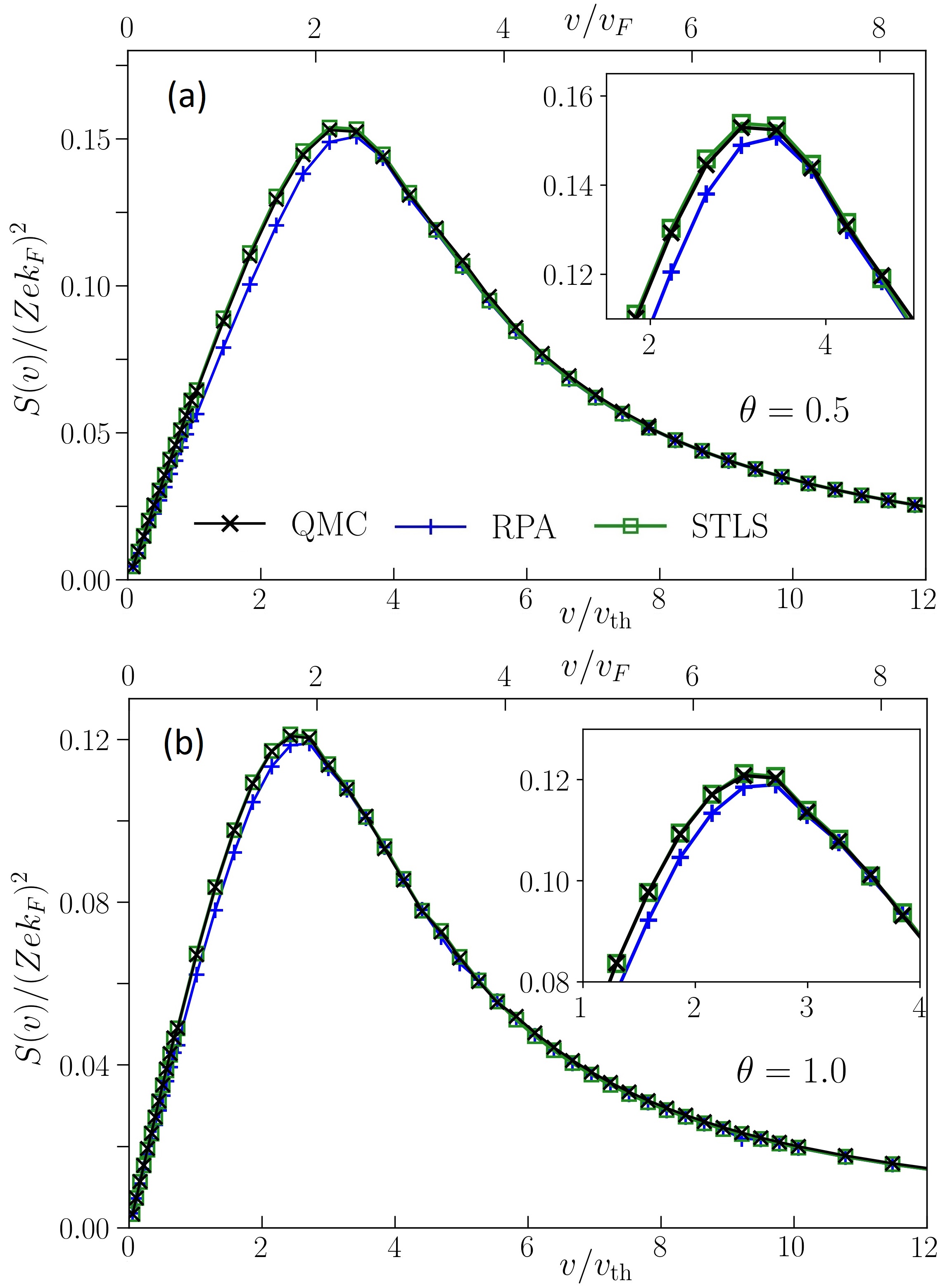}
\caption{Stopping power at $r_s=1$. Results are presented for (a) $\theta=0.5$ and (b) $\theta=1.0$. }
\label{fig:8}
\end{figure}

  \begin{figure}[h!]
 % \vspace{0.6cm}
\includegraphics[width=0.45\textwidth]{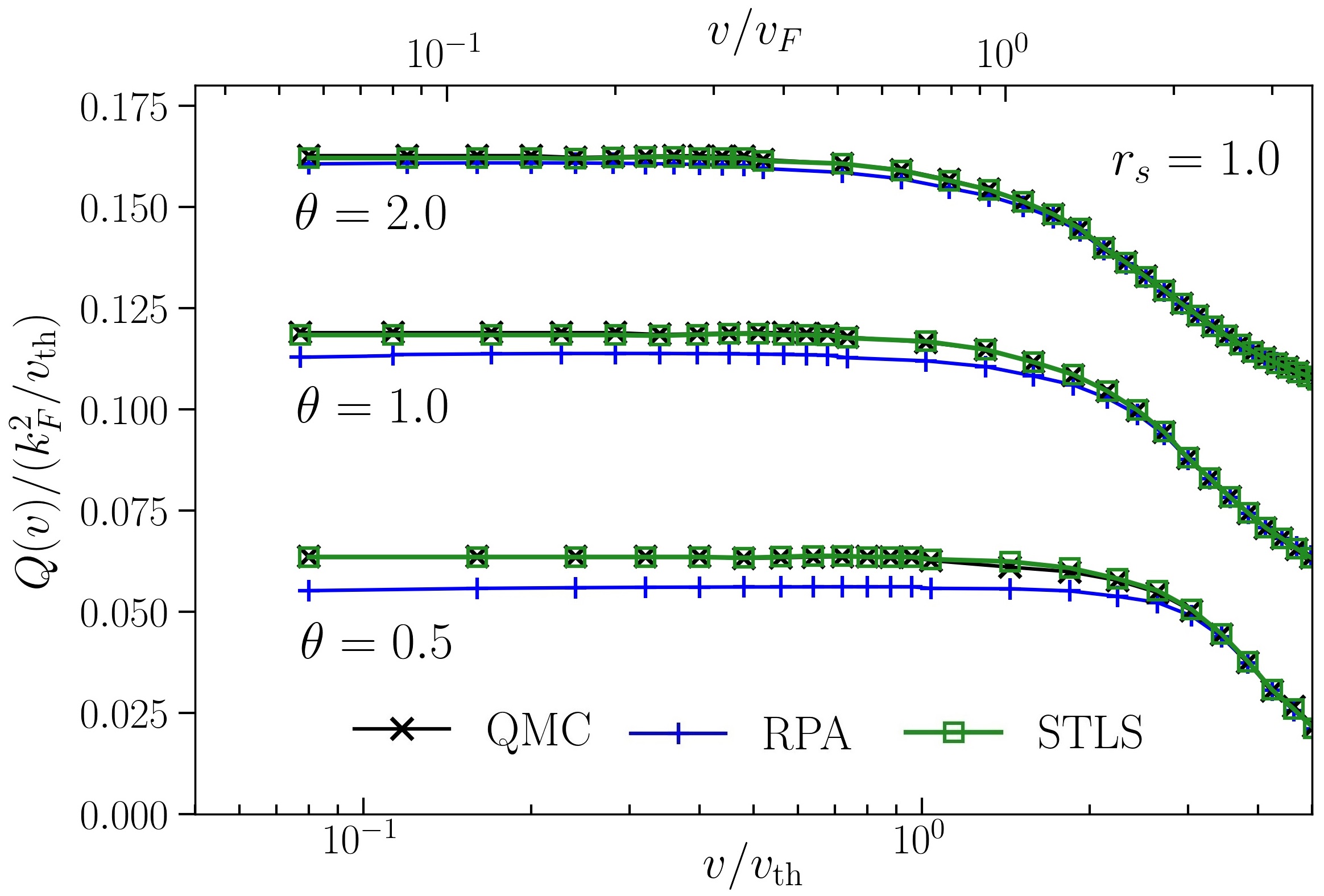}
\caption{Friction function at $r_s=1$ . Results are presented for $\theta=0.5$, $\theta=1.0$, and $\theta=2$. Curves of the friction function for different $\theta$ are shifted vertically for clarity.}
\label{fig:9}
\end{figure}

  \begin{figure}[h!]
 % \vspace{0.6cm}
\includegraphics[width=0.45\textwidth]{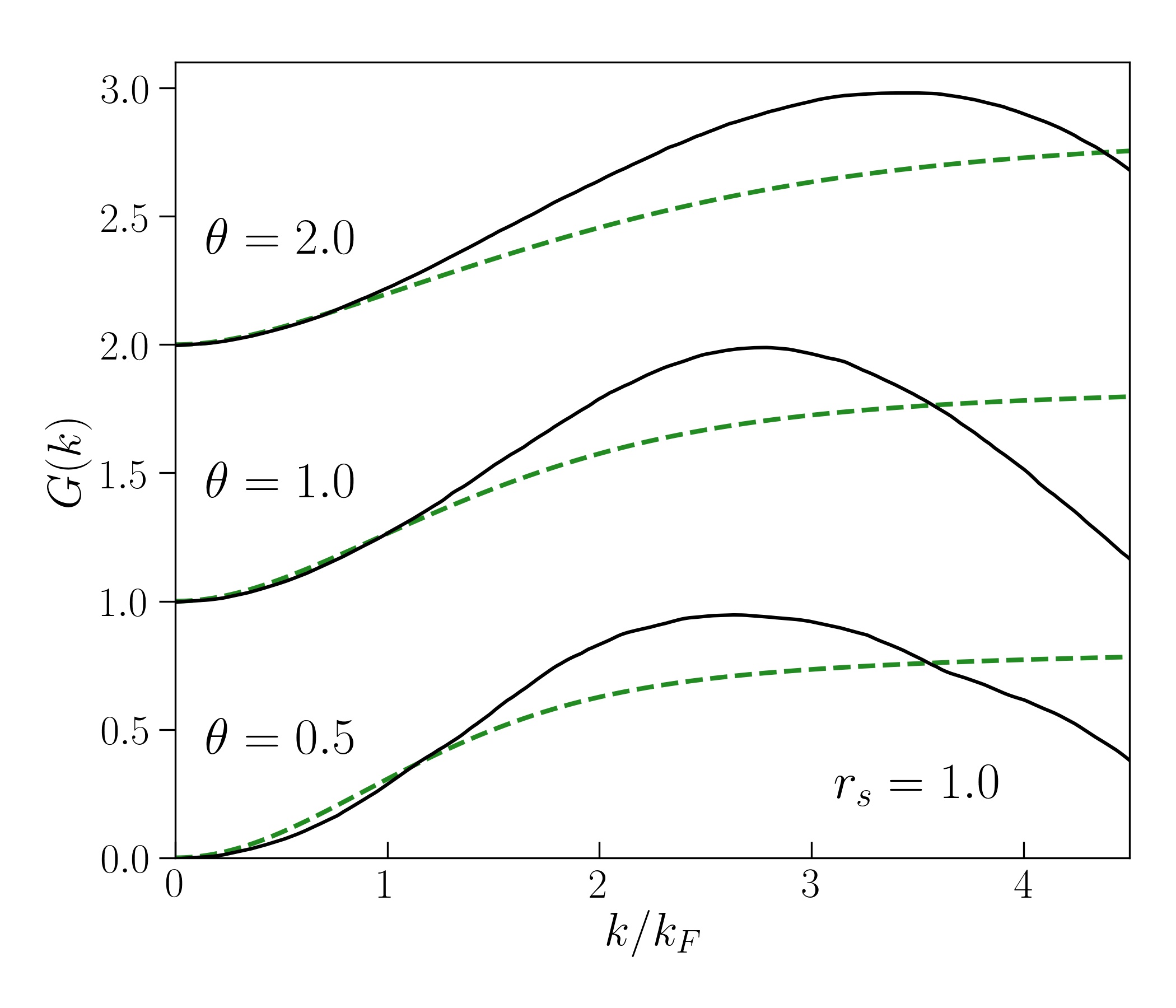}
\caption{Static local field correction at $r_s=1$ and different $\theta$. The STLS result and QMC based machine learning representation \cite{dornheim_jcp_19-nn} are shown.}
\label{fig:10}
\end{figure}

At $\theta=0.5$ and $\theta=1.0$, the stopping power and friction function are presented in Figs.~\ref{fig:8} and \ref{fig:9}. 
From Fig.~\ref{fig:8}, we clearly see that, at $r_s=1$, the effect of the electronic correlations is  significantly reduced compared to $r_s=2$ and $r_s=4$ cases due to  weaker coupling.
In contrast, but just as in the previously considered cases of $r_s=2,4$, electronic correlations do have a strong impact on the friction function. At $\theta=0.5$, the small velocity limit of the friction function increases by $20\%$ due to the electronic non-ideality. This number drops to $5\%$ at $\theta=1$ and, at $\theta=2.0$, the electronic correlations can be safely neglected, i.e.,  RPA provides an accurate description of the friction at $\theta\geq 2.0$.

Additionally, one can note from Fig.~\ref{fig:8} that the STLS results are in excellent agreement with the QMC based data, i.e., the relative agreement between the STLS and QMC results is even better than at $r_s=2$ and $r_s=4$. The same is true for the friction function (see Fig.~\ref{fig:9}). 
We compare the static LFC from STLS and from QMC in Fig.~\ref{fig:10} for $r_s=1$ and different $\theta$.
From this figure, we clearly see that the agreement between STLS and QMC results for the static LFC has been improved  at small wave numbers $k/k_F<1.5$ and decreases with $r_s$ from $r_s=2$ ($r_s=4$) to $r_s=1$, but, at larger wave numbers, the difference between  STLS and QMC data remains significant. 
Taking into account the results for $r_s=2$ and $r_s=4$, this means that, for such energy loss characteristics as the stopping power and friction function, the accurate description of the static LFC at small but finite wave numbers $k/k_F\lesssim 1$ is decisive.

  \begin{figure}[h!]
 % \vspace{0.6cm}
\includegraphics[width=0.45\textwidth]{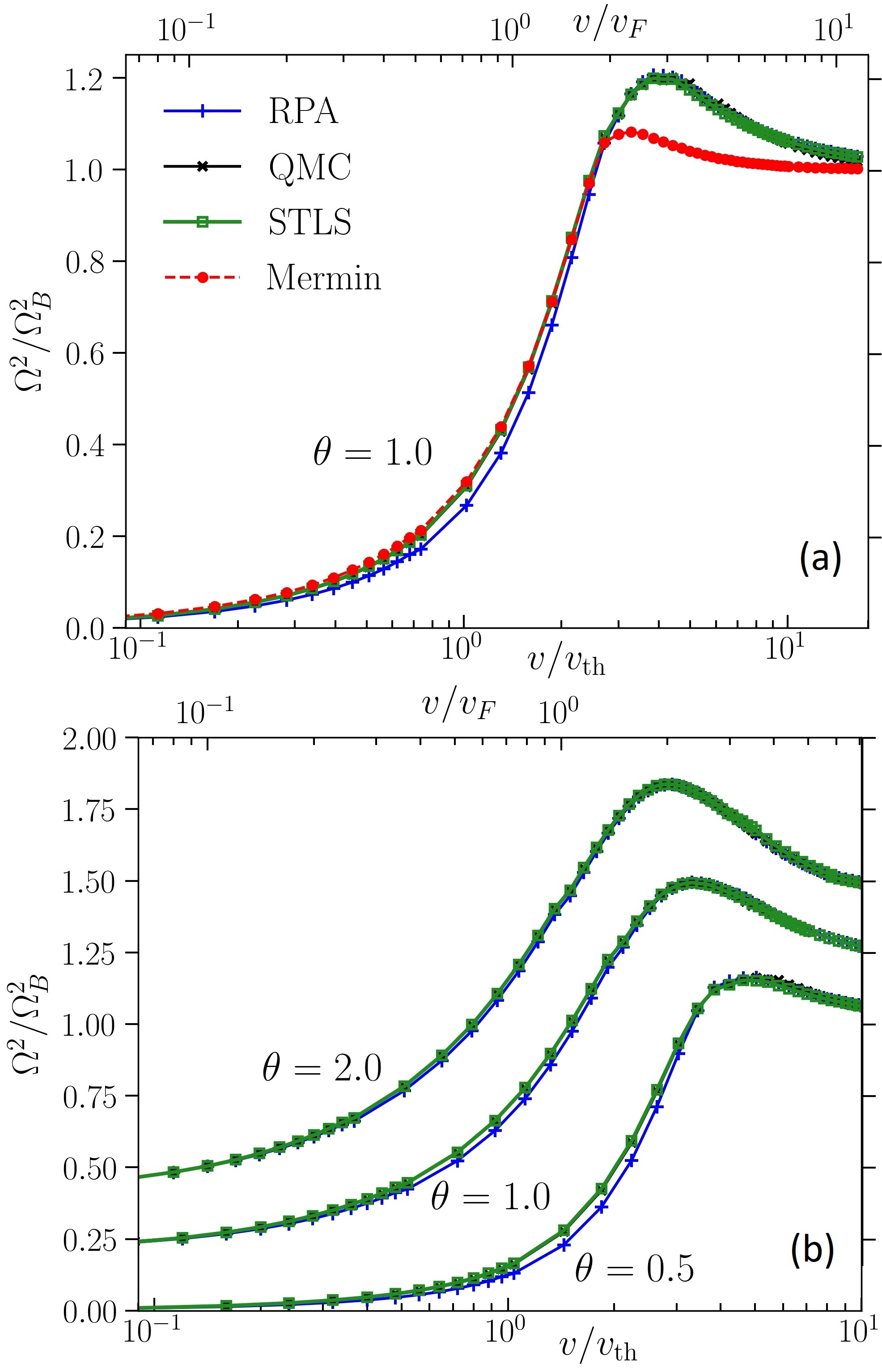}
\caption{Straggling rate at $r_s=2$. In top subplot (a), the straggling rate for $\theta=1.0$ is given. For other different values of $\theta$, the straggling rate is shown in subplot (b), where curves  for different $\theta$ are shifted vertically for clarity.}
\label{fig:11}
\end{figure}

  \begin{figure}[h!]
 % \vspace{0.6cm}
\includegraphics[width=0.45\textwidth]{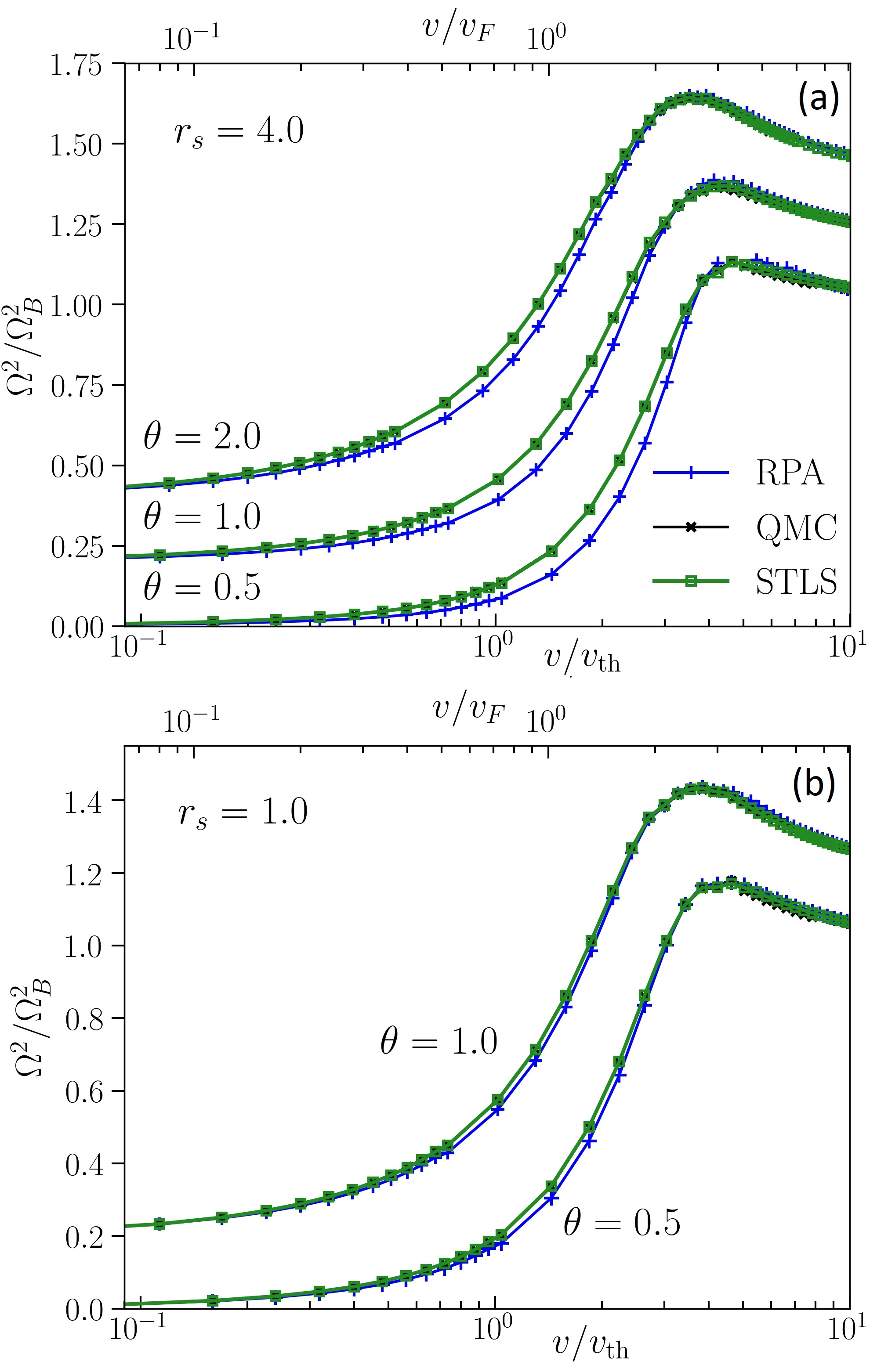}
\caption{Straggling rate for different values of the degeneracy parameter at (a) $r_s=4$ and (b) $r_s=1$. Curves  for different $\theta$ are shifted vertically for clarity.}
\label{fig:12}
\end{figure}

\subsection{Straggling rate}

To get a more complete picture on the energy loss properties of free electrons at WDM conditions, we consider next the straggling rate. 
To this end, we show the straggling rate in units of $\Omega_B=4\pi n_e Z^2$ (Bohr value for the straggling \cite{PhysRevA.23.1898}) at $r_s=2$ and different values of $\theta$ in Fig.~\ref{fig:11}.  Similarly to the stopping power, the STLS result is in agreement with the QMC data for all considered $\theta$.
%due to close values of corresponding local field corrections at $k/k_F\lesssim 1.5$.
At $\theta=0.5$ and $\theta=1.0$, the electronic correlations lead to an increase in the straggling rate by about $20\%$ at $1\leq v/v_{\rm tm}\leq 2$ and the difference between the QMC and RPA results rapidly vanishes outside of this range of projectile velocities. 
As usual, upon increasing the degeneracy parameter the effect of the correlations reduces and becomes negligible for $\theta>2.0$.  
Note that the QMC, STLS and RPA results exhibit the correct asymptotic Bohr value at large velocities, i.e., $\Omega\to\Omega_B$ at $v\gg v_{\rm th}$. 

In Fig.~\ref{fig:11}(a), we also compare the Mermin dielectric function result with $\nu_{ee}/\omega_p=0.22$ to QMC, STLS and RPA. 
Firs of all, from Fig.~\ref{fig:11}(a) we find that the Mermin result is in very good agreement with QMC and STLS at $v/v_{\rm th}<3$. However, at $3<v/v_{\rm th}<10$ the Mermin result substantially underestimates the straggling rate and tends to the Bohr limit at larger velocities. This behavior at $\theta\sim 1$ of the Mermin straggling rate 
compared to that obtained using RPA is quite general and was also reported in Ref.~\cite{Barriga_POP} for other parameters.

In Fig.~\ref{fig:12}, we present the corresponding results for $r_s=4$ and $r_s=1$ at different values of the degeneracy parameter $\theta$. Evidently, the qualitative behavior of the strangling rate as a function of velocity at these parameters is similar to the previous case of $r_s=2$. 
At $r_s=4$, as one may expect, including a static LFC has a stronger impact on $\Omega$.
From Fig.~\ref{fig:12}(a) it is clear that at $r_s=4$ the electronic correlations cannot be neglected for $0.5\leq \theta\leq 2$. Still, we note that the STLS results remain in close agreement with the QMC results due to discussed agreement of the corresponding static local field corrections (cf. Fig~\ref{fig:7}).  
At $r_s=1$, the RPA provides more accurate description of the straggling rate and electronic correlations are less significant (see Fig.~\ref{fig:12}(b)). 

\section{Implications for the Langevin dynamics of ions} \label{s:friction}

Let us now return back to the friction function and discuss in more detail the implications of our results for the  Langevin dynamics of ions.
In fact, the electronic friction can be used as a correction to the Born-Oppenheimer approximation when both ions and electrons are simulated~\cite{Wenjie}.
%The inclusion of the electronic friction in the simulation of the system of electrons and ions is, in fact, a correction to the Born-Oppenheimer approximation \cite{Wenjie}.
More specifically, within the standard Born-Oppenheimer approximation the electronic force acting on the ions is computed under the simplifying assumption that the ions are seen as immobile from the perspective of electrons.
This is justified by the large mass difference, i.e., the ions are screened statically \cite{moldabekov_pop15, zhandos_cpp17, moldabekov_pre_18, zhandos_pre_19}.
However, the motion of the ion leads to a deformation of the screening cloud around the ion, and deviations from the simple spherical symmetry appear\cite{PhysRevA.58.357, moldabekov_pre15, zhandos_cpp_19, moldabekov_cpp15, zhandos_cpp16}. 
This asymmetry in the polarization of the electrons around the ion leads to an additional electron drag force, which can be understood as a friction force. 
Indeed, from the presented results we see that at relatively small  velocities $v/v_{\rm th}\lesssim 1$, the polarization induced friction function is constant (see Figs.~\ref{fig:1}(b), \ref{fig:3}, \ref{fig:6}, and \ref{fig:9}).
This means that at these velocities the friction force acting on the ion is linearly proportional to the ion velocity. This, in turn, allows one to 
use standard Langevin dynamics for the ions to go beyond the Born-Oppenheimer approximation. Here it is important to note that the thermal velocity of the electrons is much larger than the thermal velocity of ions, since we have $v_{\rm th}=(T_e/T_i)^{1/2}(M/m_e)^{1/2}v_{\rm th}^i$,  where $v_{\rm th}^i$ is the thermal velocity of ions and $M$ is the ion mass. Therefore, the condition $v/v_{\rm th}\lesssim 1$ is equivalent to $v/v_{\rm th}^{i}\lesssim (T_e/T_i)^{1/2}(M/m_e)^{1/2}$. Taking into account that $M\gg m_e$, we conclude that, if the ions are in equilibrium with a Maxwell distribution for their velocities, the friction force is linearly proportional to the velocity for the overwhelming majority of ions. Nevertheless, there is always a small fraction of ions with $v/v_{\rm th}> 1$ ($v/v_{\rm th}^{i}\gg 1$) for which standard Langevin dynamics is not applicable. For a correct evaluation of the applicability of the Langevin approach, the number of such ions should be tracked during simulation.

The Langevin equation of motion for the ion trajectory reads \cite{Thijssen}: 
\begin{equation}\label{eq:langevin}
   M\ddot {\vec{r}}_i= \sum_{j\neq i}\vec{F}_{ij}-\gamma M\dot {\vec{r}}_i+\vec{f}_i(t),
    \end{equation}
where  $\vec{F}_{ij}$ is the force of ion $i$ on ion $j$ (taking into account the static polarization of the electronic medium), $\gamma$ the friction coefficient, and  $\vec{f}_i(t)$ is a Gaussian random force term \cite{Thijssen}. 
%Although the friction coefficient is often denoted as $\nu$, 
Here we denote the friction coefficient as $\gamma$ to avoid possible confusion with the collision frequency in the Mermin dielectric function (\ref{Mermin}).

Recalling that the stopping power can be understood as the work by a friction force, i.e. $S=\delta E/\delta l=\gamma M|\dot {\vec{r}}|$, 
we find that the friction function defined by relation Eq.~(\ref{eq:fric}) is related to the friction coefficient $\gamma$ by:
    \begin{equation}\label{eq:fric_2}
    \gamma (\theta, r_s)=\frac{Z^2e^2}{M} Q(\theta, r_s,v)\big |_{v/v_{\rm th}\ll1},
    \end{equation}
where we used the modulus of the ion velocity $|\dot {\vec{r}}|=v$.

The quantity needed for the Langevin dynamics of ions is $\gamma/\omega_{pi}$, where $\omega_{pi}$ is the plasma frequency of the ions.
To evaluate $\gamma/\omega_{pi}$ for WDM, we consider a two-component system of ions and electrons with $n_e=Zn_i$, and recast $\gamma/\omega_{pi}$ into the form:
\begin{equation}\label{eq:gamma}
  \frac{\gamma}{\omega_{pi}}=5\times10^{-2} ~\Gamma^{1/2} \left(\frac{T_i}{T_e}\right)^{1/2}\frac{Z^{2/3}}{A^{1/2}}~Q^*(\theta, r_s),  
\end{equation}
where $A$ is the mass number of an ion (atom) and $Q^*(\theta, r_s)=Q/(k_F^2/v_{\rm th})=const$ is the dimensionless value of the friction function at $v\ll v_{\rm th}$.

Using the data for the QMC friction function presented in Figs.~\ref{fig:1}(b), \ref{fig:3}, \ref{fig:6}, and \ref{fig:9},  we compiled $Q^*(\theta, r_s)$ values for different $r_s$ and $\theta$ in Table I. 
These data show that, at $0.5\leq \theta\leq 2.0$, the friction function exhibits a rather weak dependence on $\theta$.
For all considered $r_s$ values, we note that, at fixed $r_s$ (density), the value of $Q^*(\theta, r_s)$---and equivalently of $\gamma$---first increases with increasing degeneracy parameter (temperature) from $\theta=0.5$ to $\theta=1.0$, but then decreases with further increase of the degeneracy parameter to $\theta=2.0$.
This non-monotonic dependence of the friction $Q^*(\theta, r_s)$ is the result of the interplay between Pauli
blocking and thermal excitations and can be understood as follows:
At low temperatures ($\theta\ll1$), most of the electrons occupy energy levels below the Fermi energy and are not excited (scattered) by the relatively slowly moving ions~\cite{PhysRev.170.306}.
Upon increasing the temperature
%, with increase in temperature 
(so that $\theta\sim 1$), more electrons are excited to energies outside the Fermi sphere and, therefore, contribute to the scattering process, which, in turn, leads to an increasing friction.  A further increase of the temperature to weak degeneracy ($\theta>1$) makes the electrons less correlated, so that friction decreases again. 
%and, thus, leads to the decrease of the friction.

Using the QMC based data for $Q^*(\theta, r_s)$ and Eq.~(\ref{eq:gamma}), we can now estimate $\gamma/\omega_{pi}$ for experimentally realized WDM conditions and dense plasma parameters. As an example, we take parameters of a non-ideal plasma from (i) the direct-drive inertial-confinement-fusion (ICF) experiments on OMEGA \cite{Hu_2010, BOEHLY1997495} and (ii) parameters of WDM from experiments on solid Be heated by $(4-5) ~{\rm keV}$ pump photons \cite{LANDEN2001465}.
For the case (i),  considering a hydrogen plasma with $T_e/T_i=1$, $\Gamma=6$, $r_s=2$ and $\theta=0.5$ we find $\gamma/\omega_{pi}\simeq 0.012$. For the case (ii), we find  $\gamma/\omega_{pi}\simeq 0.01$ for beryllium at WDM conditions with $T_e/T_i=1$, $Z=2$, $\Gamma=10$, $r_s=2$ and $\theta=1.0$.

\begin{table}[h]
      \centering
          \caption{Values of the normalized friction function, $Q^*(\theta, r_s)$, for different values of $r_s$ and $\theta$.}\label{tab:1}
      \vspace{0.5cm}
%      \begin{tabular}{|c| c| c| c|}\hline
  %    \\[-0.5em]
%          $r_s$ &  $\theta=0.5$ &  $\theta=1.0$ & $\theta=2.0$ \\
%          \hline\hline
%        4.0& 0.154 & 0.178 & 0.172\\\hline
%         2.0& 0.0989 & 0.11 & 0.105\\\hline
%            1.0& 0.0636 & 0.07 & 0.0626\\\hline
%      \end{tabular}
      \begin{tabular}{|c| c| c| c| c |}\hline
  %    \\[-0.5em]
          & $\theta$ &$0.5$ &  $1.0$ & $2.0$ \\
%          \hline\hline
          $r_s$ &  & & &  \\
          \hline\hline
        4.0&& 0.1540 & 0.178 & 0.1720\\\hline
         2.0&& 0.0989 & 0.110 & 0.1050\\\hline
            1.0&& 0.0636 & 0.070 & 0.0626\\\hline
      \end{tabular}
      \end{table}

We note that for the presented examples, the previously used Rayleigh model for the friction by free electrons \cite{PhysRevLett.104.245001, Kang_2018}, $\gamma=2\pi Z (m_e/M) (v_{\rm th}/a_i)$, overestimates the friction coefficient, compared to our new QMC-based results, by about three and two times for the cases (i) and (ii), respectively.

To get a picture about the importance of friction with $\gamma/\omega_{pi}\approx 0.01$, we refer to a recent analysis of the dynamical structure factor of a one-component Yuakwa system  by K\"ahlert \cite{Hanno_POP}, where it was shown that, at $\Gamma=10$, a friction coefficient $\gamma/\omega_{pi}=0.01$ is sufficient to decrease the dynamical structure factor peak height at $ka_i=0.25$  by a factor two. In general, at $\gamma/\omega_{pi}\approx 0.01$, sound waves of the non-ideal ions are highly sensitive to friction \cite{Hanno_POP, Mabey, PhysRevE.79.046412}. 

\section{Summary} \label{s:dis}
First of all, the comparison of the QMC based results to RPA allowed us to gauge the importance of electronic correlations at typical WDM conditions. 
In particular, we provide a quantitative analysis of electronic correlation effects by considering partially degenerate electrons with $1\leq r_s \leq 4$. The electronic non-ideality increases the stopping power (and straggling rate) up to about its maximum, but has no impact at larger velocities. (Interestingly, the same effect of electronic correlations has recently been observed for ion stopping in correlated solids \cite{balzer_prb16}).
At $r_s=1$, the account of correlation effects has only a weak impact, which indicates that for $r_s<1$---in contrast to the $r_s>1$ case---the electronic correlations are not important for an accurate description of energy loss characteristics. 

Secondly, we compared our new QMC-based results to those obtained using the Mermin dielectric function. 
The corresponding analysis shows that the applicability of the relaxation time approximation for the description of the stopping power and straggling rate at WDM conditions is restricted to low velocities $v\lesssim v_{\rm th}$. At larger velocities, in addition to quantitative differences, the qualitative behavior of the  Mermin stopping power and straggling rate show significant discrepancies to the QMC based results. Although the Mermin result for the stopping power can be in a good agreement with the QMC based data at $v\approx v_{\rm th}$, the corresponding  friction functions at $v\ll v_{\rm th}$ exhibit substantial differences. Moreover, adjusting the collision frequency to fix the latter disagreement for friction function values results in a worsening of the performance of the Mermin approach for the stopping power at $v\approx v_{\rm th}$. In a nutshell, the applicability of the relaxation time approximation for the description of the energy loss properties of correlated partially degenerate electrons is severely limited.   

Thirdly, we analysed the degree of approximation of the static LFC needed to adequately describe energy loss properties. This was done by comparing the data based on the QMC results for the static LFC to that computed using STLS. At $1\leq r_s\leq 4$ and $0.5\leq \theta\leq 4$, the STLS static local field correction based data for the considered energy loss properties is in remarkably good agreement with the results obtained using the QMC based machine learning representation of the static LFC. 
By comparing the STLS and QMC data for $G(k)$, we have found the general conclusion that accurate data for the LFC at $k/k_{F}\lesssim 1.5$ are important for a correct description of the stopping power, straggling rate and friction function. If the latter requirement is fulfilled, even a significant disagreement at larger wave numbers between the STLS static LFC with that from QMC simulations is not critical.   
In other words, the considered energy loss  properties are not sensitive to the inaccuracy of the approximation for the static LFC at larger wave numbers, $k/k_{F}>2$ at the considered parameters.  

Finally, we analyzed the friction function and the related friction coefficient of electrons at WDM conditions. In particular, the electronic friction is of paramount importance for the adequate Langevin dynamics simulation of ions.  Compared to the stopping power and straggling rate, the friction function (coefficient) is much more sensitive to the inclusion of the electronic correlations for relatively low velocity of ions $v\lesssim v_{\rm Th}$.  The electronic correlations result in a significant increase of the friction coefficient compared to RPA and cannot be neglected at $1\leq r_s \leq 4$ with $\theta \sim 1$. Moreover, we found that the previously used Rayleigh  model for electronic friction ~\cite{PhysRevLett.104.245001, Kang_2018} significantly overestimates the friction coefficient and is not applicable for WDM and non-ideal dense plasma conditions. For experimentally obtained WDM and dense plasmas with $r_s=2$ and $\theta\sim 1$, the friction function is about $\gamma/\omega_{pi}\approx 0.01$. 
Therefore, as discussed in Sec.~\ref{s:friction},  dissipation effects beyond Born-Oppenheimer approximation are important for an adequate simulation of the dynamical properties of non-ideal ions.

 \section*{Acknowledgments}
   This work has been supported by the Ministry of Education and Science of Kazakhstan under Grant No. BR05236730, “Investigation of fundamental problems of
physics of plasmas and plasma-like media” (2020) and by the Norddeutscher Verbund f\"ur Hoch- und H\"ochstleistungsrechnen (HLRN) via computation time under grant \emph{shp00015}. MB acknowledges support by the Deutsche Forschungsgemeinschaft via grants BO1366/13 and BO1366/15. 
TD acknowledges support by the Center for Advanced Systems Understanding (CASUS) which is financed by Germany's Federal Ministry of Education and Research (BMBF) and by the Saxon Ministry for Science and Art (SMWK) with tax funds on the basis of the budget approved by the Saxon State Parliament.
Fruitful discussions with H. K\"ahlert are gratefully acknowledged.

\end{document}